\documentclass{article}
\usepackage{spconf,amsmath,graphicx,hyperref}

\usepackage{booktabs}
\usepackage{multirow}
\usepackage{amsmath}
\usepackage{cite}
\usepackage{wasysym}
\usepackage{xcolor}

\newcommand{\Fig}[1]{\textbf{Fig.~\ref{fig:#1}}} % refer to figure
\newcommand{\Table}[1]{\textbf{Table~\ref{tb:#1}}} % refer to table
\newcommand{\Sec}[1]{\textbf{Section~\ref{sec:#1}}} % refer to section
\newcommand{\Subsec}[1]{\textbf{Section~\ref{subsec:#1}}} % refer to subsection

\newcommand{\Argmax}{\mathop {{\rm argmax}}}

\title{Voting-based Pitch Estimation with Temporal and Frequential Alignment and Correlation Aware Selection}
\name{Junya Koguchi, Tomoki Koriyama}
\address{
CyberAgent, Inc. \\
\{koguchi\_junya, koriyama\_tomoki\}@cyberagent.co.jp
}

\begin{document}
\ninept
\maketitle
\begin{abstract}
The voting method, an ensemble approach for fundamental frequency estimation, is empirically known for its robustness but lacks thorough investigation. This paper provides a principled analysis and improvement of this technique. First, we offer a theoretical basis for its effectiveness, explaining the error variance reduction for fundamental frequency estimation and invoking Condorcet's jury theorem for voiced/unvoiced detection accuracy. To address its practical limitations, we propose two key improvements: 1) a pre-voting alignment procedure to correct temporal and frequential biases among estimators, and 2) a greedy algorithm to select a compact yet effective subset of estimators based on error correlation. Experiments on a diverse dataset of speech, singing, and music show that our proposed method with alignment outperforms individual state-of-the-art estimators in clean conditions and maintains robust voiced/unvoiced detection in noisy environments.
\end{abstract}
\section{Introduction}
\label{sec:intro}
The fundamental frequency ($f_\mathrm{o}$~\footnote{We use $f_\mathrm{o}$ as recommended by Acoustical Society of America. The notation F0 is misleading since it may suggest the “0th formant,” and it can be confused either with the lowest resonance frequency of audio devices or with the lowest musical note F.}) is a physical quantity that roughly corresponds to pitch, and it is widely used in speech and music information processing.
$f_\mathrm{o}$ is defined as the reciprocal of the shortest period of a periodic signal.
$f_\mathrm{o}$ estimation methods that focus on this periodic structure include those that use extrema of the waveform autocorrelation function or zero crossings~\cite{Mauch2014pyin, morise2009dio, talkin2015reaper, morise2017harvest}.
For speech and some musical tones, one can assume an integer-multiple harmonic structure.
Based on this assumption, methods have been proposed that use the spacing between harmonics or the agreement with a harmonic pattern~\cite{Camacho2008swipe, miwa2017freedam}.
In contrast to such deterministic approaches, statistical $f_\mathrm{o}$ estimators in which a deep neural network (DNN) replaces manual signal model design and ad~hoc processing outperform deterministic methods in noise robustness and estimation accuracy~\cite{Kim2018crepe}.
Following this success, improvements have been proposed that include architectural design~\cite{Li2024yolopitch} and the use of self-supervised learning~\cite{Gfeller2020spice, Riou2023pesto}.
However, both deterministic and statistical methods are vulnerable to signals that deviate from the assumptions made during algorithm design or from the domains covered by the training data~\cite{morrison2023penn}.

\begin{figure}[tb]
    \centering
    \includegraphics[width=\linewidth]{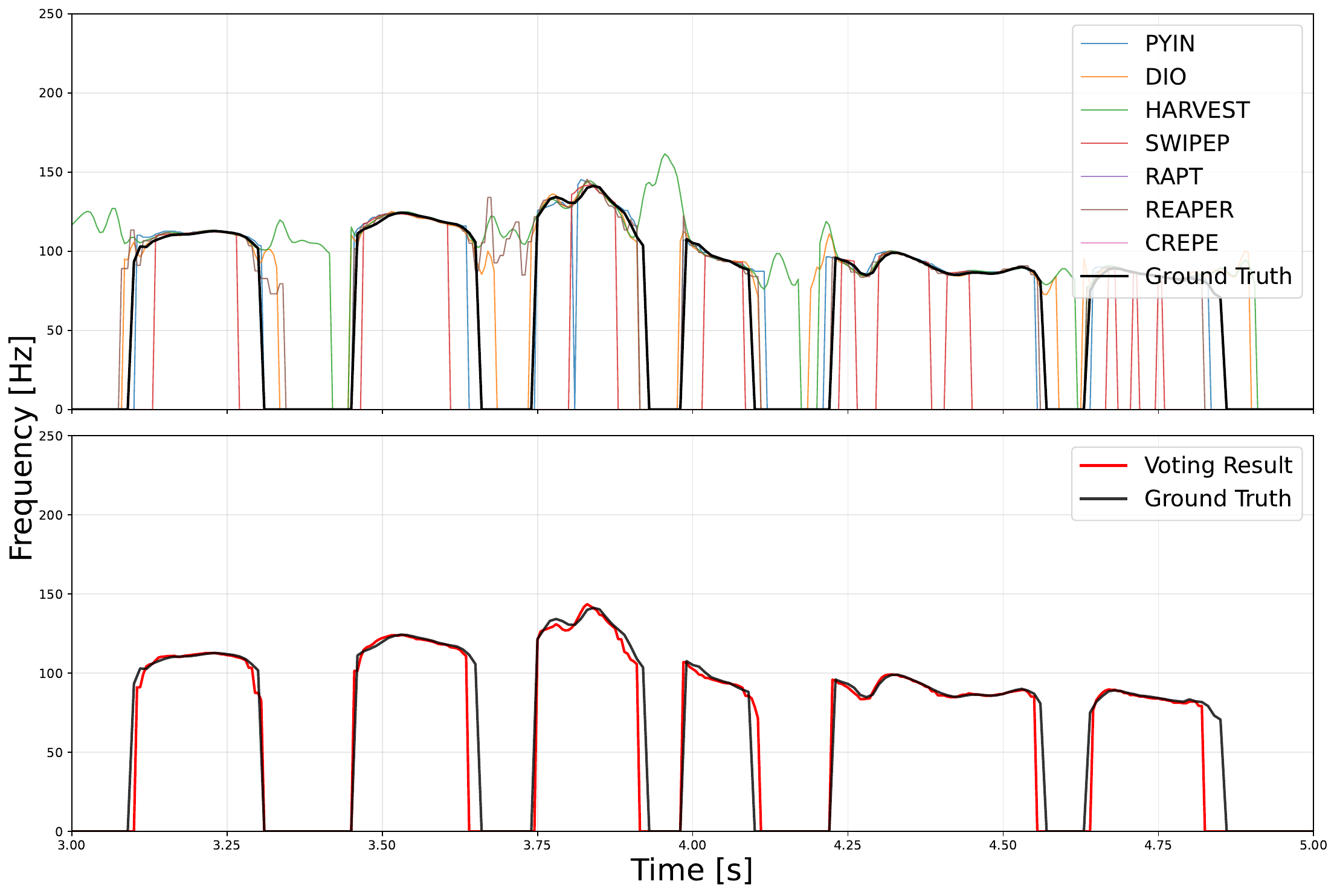}
    \vspace{-20pt}
    \caption{Comparison between a single method and the voting method.
    Voting absorbs estimation errors and produces robust estimates.}
    \label{fig:voting}
\end{figure}

On the other hand, there is a voting method that estimates $f_\mathrm{o}$ by aggregating the outputs of multiple methods through voting~\cite{Drugman2018ensemble}.
This strategy is empirically known to be effective in speech synthesis~\cite{Yamagishi2008voting, Qian2009voting, Liu2021diffsvc, Ohtani2025firnet}.
It is widely used because it can absorb estimation errors by majority decision, it tends to be robust, and it is easy to implement (\Fig{voting}).
Despite its practical use, investigations of the voting method are limited and several issues remain.
1) There is insufficient discussion of why it improves accuracy over a single method and how each combination contributes to the improvement.
2) Due to differences such as the choice of center time in the short-time Fourier transform (STFT), time shifts can arise between the estimated $f_\mathrm{o}$ sequences produced by different methods for the same signal~\cite{Wang2024alignment}.
If we aggregate such sequences without correction, we may oversmooth temporal variations or degrade accuracy near voiced or unvoiced (V/UV) boundaries.
3) Theoretical guarantees for accuracy improvement under voting require low correlation among estimation errors.
With a small number of methods, an ensemble of strongly correlated methods that cause error at the same locations can degrade accuracy.

This work studies, improves, and evaluates the voting method, which has not been sufficiently examined.
First, we analyze why accuracy improves by aggregation to the median, from the viewpoints of variance reduction of errors and the convergence of votes to the correct value.
Based on these considerations, we propose two improvements.
1) We correct for method-dependent biases by aligning the estimated sequences in time and frequency with a reference.
2) We select a small set of methods whose corrected estimation errors show low pairwise correlation, which enables high accuracy with fewer methods.
Finally, we provide an experimental evaluation and discuss results based on several error measures and on the quality of analysis-by-synthesis audio.

\section{Theoretical considerations of the voting method}
\label{sec:voting}
\subsection{Related work}
The voting method aggregates multiple estimates as follows.
Within voiced frames it uses the median of the estimated frequencies (the mean of the two central values when the number of methods is even), and for V/UV decisions it uses the mode.
When combining estimators that rely on different acoustic features through various ensemble schemes, median aggregation has been shown experimentally to yield the highest accuracy~\cite{Drugman2018ensemble}.
Its effectiveness is also known empirically in speech synthesis research, and it achieves high accuracy for both the estimated frequency trajectory and V/UV flags~\cite{Yamagishi2008voting, Qian2009voting, Liu2021diffsvc, Ohtani2025firnet}.

\subsection{Variance reduction}
\label{subsec:variance}
Consider aggregating the outputs of $n$ methods.
Let $\theta$ denote the ground-truth value for a single frame of a signal (the true $f_\mathrm{o}$), and let $X_i$ denote the estimate from the $i$-th method.
With the estimation error $\varepsilon_i$, we write
\begin{equation}
    X_i = \theta + \varepsilon_i, \quad i=1, \dots, n .
\end{equation}
Assume that $\mathrm{median}(\varepsilon_i)=0$ and that the average correlation between the error signs $\psi_i=\mathrm{sign}(\varepsilon_i)$ is less than one.
Let $\tilde{X}$ be the median of $\{X_i\}_{i=1}^n$.
The empirical distribution function $G_n(\tilde{X})$ at $\tilde{X}$ equals $1/2$:
\begin{equation}
    G_n(\tilde{X}) = \frac{1}{n}\sum_{i=1}^n{I \left( X_i < \tilde{X} \right)} = \frac{1}{2} ,
\end{equation}
where $I(\cdot)$ is the indicator function that returns $1$ if the input condition is true and $0$ otherwise.
A first-order Taylor expansion of $G_n(\tilde{X})$ around the true value $\theta$ gives
\begin{equation}
    G_n(\tilde{X}) \approx G_n(\theta) + {G_n}'(\theta)\left(\tilde{X}-\theta\right) .
\end{equation}
We can approximate $G_n(\theta)$ and ${G_n}'(\theta)$ as
\begin{align}
    G_n(\theta) = \frac{1}{n}\sum_{i=1}^n I(X_i < \theta) &= \frac{1}{n}\sum_{i=1}^n I(\varepsilon_i < 0)\notag\\
    &\approx \frac{1}{2} - \frac{1}{2n} \sum_{i=1}^n{\psi_i}, \\
    {G_n}'(\theta) \approx \frac{1}{n}\sum_{i=1}^n h_i(0) ,
\end{align}
where $h_i$ is the probability density function of $\varepsilon_i$.
Solving for $\tilde{X}-\theta$ yields
\begin{equation}
   \tilde{X} - \theta \approx \frac{1}{2\sum_{i=1}^n h_i(0)}  \sum_{i=1}^n \psi_i .
\end{equation}
Therefore, the variance of the median error is approximated by
\begin{equation}
    \mathrm{Var}\left(\tilde{X} - \theta\right) \approx \frac{1}{4\left\{\sum_{i=1}^n h_i(0)\right\}^2} \mathrm{Var}\left(\sum_{i=1}^n \psi_i\right) .
\end{equation}
Given $\mathrm{median}(\varepsilon_i) = 0$, we have $E[\psi_i]=0$ and, since $\psi_i^2 = 1$, we have $\mathrm{Var}(\psi_i) = 1$.
Let $\rho_{ij} = \mathrm{Corr}(\psi_i, \psi_j)$ be the correlation coefficient.
Then
\begin{equation}
    \mathrm{Var}\left(\sum_{i=1}^n \psi_i\right) = n + 2 \sum_{i<j}^n \rho_{ij} .
\end{equation}
Using the average correlation $\bar{\rho}= \frac{2}{n(n-1)} \sum_{i<j}^n \rho_{ij}$, we obtain
\[
    \mathrm{Var}(\tilde{X} - \theta) \approx \frac{1}{4 \left\{ \sum_{i=1}^n h_i(0) \right\}^2 } \{n+n(n-1)\bar{\rho}\} .
\]
With $\sum_{i=1}^n h_i(0) = n\bar{h}$, where $\bar{h}$ is the mean of $h_i(0)$, we finally have
\begin{equation}
    \mathrm{Var}(\tilde{X} - \theta) \approx \frac{1+(n-1)\bar{\rho}}{4n\bar{h}^2} .
\end{equation}
Since $\bar{\rho}<1$ by assumption, this result suggests that the error variance decreases as the number of methods $n$ increases, compared with using a single method ($n=1$).
As a practical example, when outliers such as octave errors occur, the median is insensitive to a few large errors as long as the majority of estimators return values near the true value.
In this sense, the median is a robust estimator.

\subsection{Condorcet's jury theorem}
The use of the mode for V/UV decisions can be explained by Condorcet's jury theorem~\cite{condorcet1785jury}.
Consider a binary classification problem that decides a V/UV flag for each STFT analysis frame.
For simplicity, assume that each method is correct with probability $p$ and that errors are independent across methods.
With $n$ methods, the probability $P_n$ that the majority vote is correct is
\begin{equation}
P_n= \sum_{m = \lceil (n+1)/2 \rceil}^{n} \binom{n}{m} p^m (1-p)^{n-m} .
\end{equation}
If $p>0.5$ and the decision errors are independent, the accuracy of the majority vote exceeds the accuracy $p$ of any single method, and $P_n$ approaches one as $n$ increases.
This property supports the expectation that using the mode among methods with independent errors improves V/UV accuracy.

\section{Improvements to the voting method}
\subsection{Correction by alignment in time and frequency}
\label{subsec:alignment}
\begin{figure}[tb]
    \centering
    \includegraphics[width=0.9\linewidth]{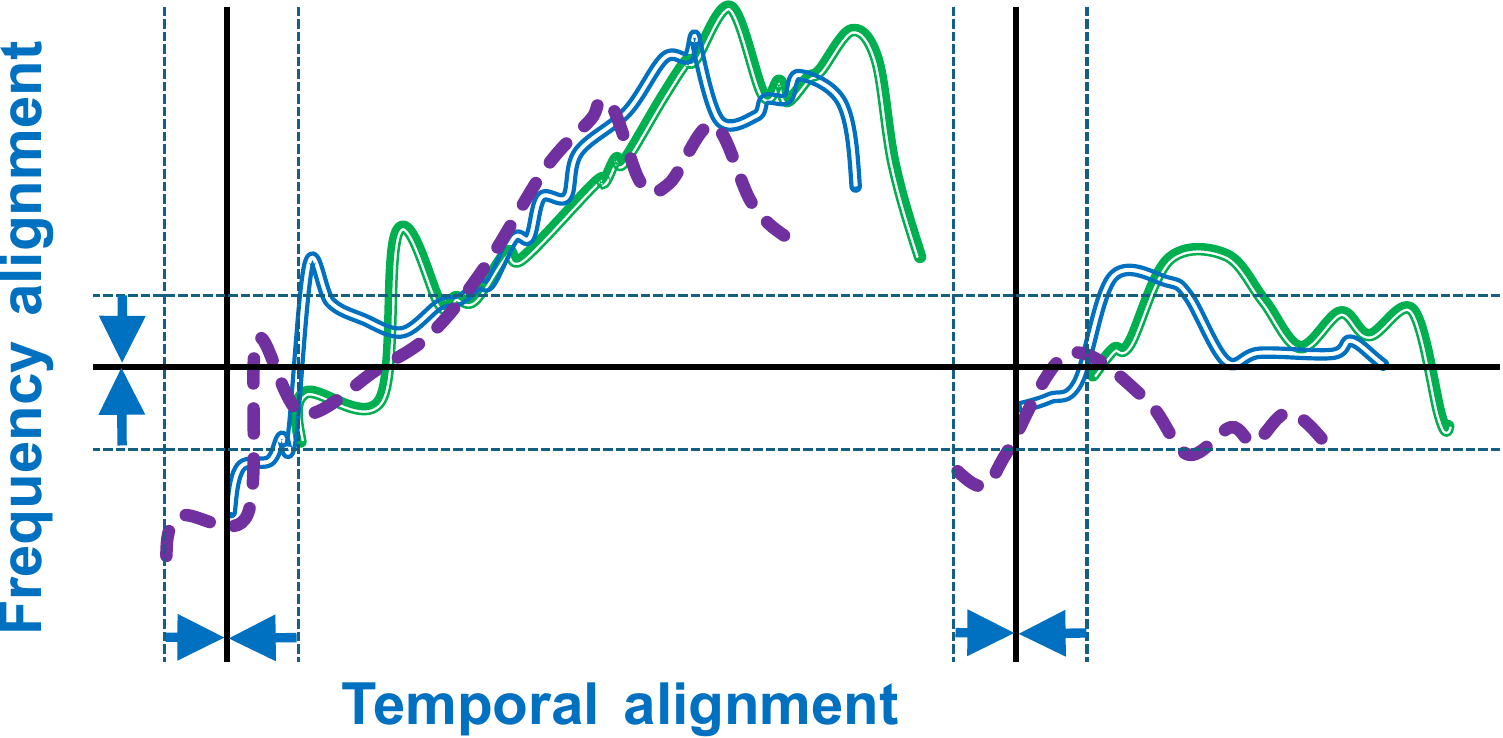}
    \vspace{-10pt}
    \caption{Correction of analysis-time offsets by alignment.}
    \label{fig:alignment}
\end{figure}
Before aggregation by voting, we propose an algorithm that corrects time and frequency offsets among estimated $f_\mathrm{o}$ sequences (\Fig{alignment}).
Differences in the analysis frame center of the STFT or in peak-picking positions can cause method-dependent biases along both the time axis and the frequency axis.
These biases break the assumptions in \Subsec{variance} and can degrade accuracy when we aggregate estimates.

For time-axis correction, we adapt a method that was originally proposed for evaluation with temporal alignment~\cite{Wang2024alignment}.
We first choose a reference method and, for each other method, we find a target offset $k_\mathrm{target}$ that maximizes similarity to the reference.
Let the analysis-frame length be $L$, the frame index be $l$, the reference sequence be $f_\mathrm{ref}$, and the estimated sequence be $\hat{f}$.
For each candidate offset $k$, we align $\hat{f}$ to $f_\mathrm{ref}$ and compute the similarity between the two pitch sequences.
We use raw pitch accuracy (RPA) as the similarity metric:
\begin{equation}
    \mathrm{RPA}_{\epsilon}(k) = \frac{1}{L }\sum_{l=0}^{L-1}
     \begin{cases}
     1, & \mathrm{if}~\Delta\cent \left( \hat{f}(l+k), f_\mathrm{ref}(l) \right) < \epsilon , \\
     0, & \mathrm{otherwise} .
     \end{cases}
\end{equation}
Here, $\Delta\cent$ denotes the interval in cents of $\hat{f}$ relative to $f_\mathrm{ref}$:
\begin{equation}
   \Delta \cent(\hat{f}, f_\mathrm{ref})=1200 \log_2 \frac{\hat{f}}{f_\mathrm{ref}} .
\end{equation}
$\mathrm{RPA}_\epsilon$ counts a frame as correct if the deviation from the reference is within $\epsilon$ cents and as incorrect otherwise.
We allow a tolerance because spectral leakage and period jitter can cause small discrepancies even when methods track the same pitch.
For each method, we obtain the alignment offset
\begin{equation}
k_\mathrm{align} = \Argmax_{-H \le k \le H} \mathrm{RPA}_\epsilon(k) ,
\end{equation}
where $H$ is the search range in frames, which bounds the maximum expected temporal offset between two $f_\mathrm{o}$ sequences.
We shift $\hat{f}$ by $k_\mathrm{align}$ according to its sign.
To maintain the sequence length $L$, we zero-pad and truncate as needed.
Since we insert a short silence at the beginning and end of the analysis signal, the effect of padding and truncation on speech segments is negligible.

We further remove frequency-axis biases.
Within voiced frames, we compute the frequency bias $f_\mathrm{align}$ as
\begin{equation}
f_\mathrm{align} = \mathrm{median}{\left( \Delta\cent \left(\hat{f}(l+k_\mathrm{align}), f_\mathrm{ref}(l)\right) \right)} .
\end{equation}
The median suppresses sharp spikes and local double or half pitch errors and aggregates only the global bias.
We compute this quantity for each method and subtract it in the cent domain to correct the estimated $f_\mathrm{o}$.

\subsection{Method selection by a greedy search}
\label{subsec:selection}
According to the discussion in \Sec{voting}, we expect accuracy to improve as the number of methods increases.
However, when computation must be reduced, it is useful to restrict the number of methods used for voting.
To support this use case, we propose a greedy algorithm that selects methods from a candidate pool.
Let $S = \{A_1, A_2, \dots, A_n\}$ denote the set of all candidate methods.
We prepare an initial selected set $G = \{A_0\}$.
For each $A_j \in S$, we tentatively add $A_j$ to $G$ and evaluate the resulting ensemble.
We use two criteria for evaluation.
1) RPA as a direct measure of estimation accuracy.
2) The average correlation of error signs as a measure that supports the assumptions in \Subsec{variance}.
We add to $G$ the method that yields the largest RPA or, under the correlation criterion, the smallest average correlation within $G$, and we remove it from $S$.
We stop when adding a new method does not improve the score or when $G$ reaches a predefined size.
The final set $G$ is used for voting.

\begin{table*}[tb]
\centering
\caption{Evaluation without additive noise. REAPER is grayed out because it was used to estimate the ground truth for speech.}
\label{tb:no_noise}
\renewcommand{\arraystretch}{0.9}
\resizebox{\textwidth}{!}{
\begin{tabular}{lcccccccc}
\toprule
\multirow{2}{*}{Method} & \multirow{2}{*}{$\Delta \cent$ $\downarrow$} & \multicolumn{3}{c}{RPA$ \uparrow$} & \multicolumn{2}{c}{V/UV} & \multirow{2}{*}{UTMOS$ \uparrow$} \\
\cmidrule(lr){3-5} \cmidrule(lr){6-7}
& & 5 & 25 & 50 & Recall $\uparrow$ & False alarm $\downarrow$ & \\
\midrule
\color{gray}{REAPER}  &  \color{gray}{3.11 $\pm$ 151.27} & \color{gray}{39.34} & \color{gray}{72.97} & \color{gray}{80.56} & \color{gray}{91.68} &  \color{gray}{8.72} & \color{gray}{1.20} \\
\cmidrule[0.05mm](lr){1-8}
RAPT    &   3.57 $\pm$ 191.78 & 26.81 & 64.11 & 75.06 & 91.65 & 13.16 & 1.20 \\
SWIPE'  & 143.35 $\pm$ 195.89 & 15.29 & 50.40 & 80.91 & 72.94 & 30.12 & 1.21 \\
pYIN    &  20.37 $\pm$ 123.77 & 21.37 & 56.54 & 72.30 & 67.86 & \textbf{16.13} & 1.21 \\
DIO     &  -9.65 $\pm$ 119.30 & 27.25 & 64.15 & 74.02 & 78.16 & 23.48 & 1.22 \\
Harvest &  32.37 $\pm$ 190.23 & 23.97 & 63.47 & 76.49 & 90.48 & 64.83 & 1.22 \\
Praat   &  41.34 $\pm$ 122.03 &  9.66 & 34.55 & 53.01 & 84.94 & 34.96 & 1.22 \\
CREPE   &  12.27 $\pm$ 106.47 & 13.07 & 50.68 & 75.84 & 87.98 & 20.50 & \textbf{1.23} \\
FCNF0++ &  25.07 $\pm$ 258.63 & 18.21 & 55.48 & 68.91 & 89.92 & 18.77 & 1.20 \\
\cmidrule[0.05mm](lr){1-8}
Voting (all methods)  &  \textbf{3.35} $\pm$ 188.20 & \textbf{29.01} & \textbf{66.89} & \textbf{76.78} & \textbf{94.21} & 19.29 & \textbf{1.23} \\
\textcolor{black}{\quad w/o frequential alignment} & 20.18 $\pm$ 191.45 & 19.46 & 66.31 & 76.78 & 94.20 & 19.29 & 1.22 \\
\textcolor{black}{\quad \quad  w/o temporal alignment} & 40.11 $\pm$ 161.03 & 22.39 & 65.32 & 76.01 & 93.99 & 20.70 & 1.20  \\
\bottomrule
\end{tabular}
}
\end{table*}

\begin{table}[tb]
\centering
\caption{RPA under additive noise (threshold 50 cents). REAPER is grayed out because it was used to compute the speech ground truth.}
\vspace{5pt}
\label{tb:noise_rpa}
\renewcommand{\arraystretch}{1}
\resizebox{\columnwidth}{!}{
\begin{tabular}{lccccc}
\hline
\multirow{2}{*}{Method $\backslash$ SNR [dB]} & \multicolumn{5}{c}{$\mathrm{RPA}_{50}$$ \uparrow$} \\
\cmidrule(lr){2-6}
& $\infty$ & 30 & 20 & 10 & 0 \\
\hline
\color{gray}{REAPER}  & \color{gray}{80.56} & \color{gray}{80.24} & \color{gray}{78.01} & \color{gray}{68.85} & \color{gray}{37.45} \\
\cmidrule[0.05mm](lr){1-6}
RAPT    & 75.06 & 57.24 & 56.71 & 53.54 & 36.85 \\
SWIPE'  & 80.91 & 60.44 & 58.49 & 49.01 & 18.35 \\
pYIN    & 72.30 & 71.74 & 65.39 & 61.85 & 26.27 \\
DIO     & 74.02 & 55.68 & 51.90 & 42.68 & 17.18 \\
Harvest & 76.49 & 67.60 & 66.75 & 57.54 & 25.51 \\
Praat   & 53.01 & 31.95 & 31.65 & 28.16 & 13.66 \\
CREPE   & 75.84 & 64.69 & 64.38 & \textbf{62.31} & \textbf{50.65} \\
FCNF0++ & 68.91 & 70.32 & \textbf{68.80 }& 56.52 & 22.86 \\
\cmidrule[0.05mm](lr){1-6}
Voting  & \textbf{76.78} & \textbf{71.90} & 60.40 & 61.50 & 42.27 \\
\hline
\end{tabular}
}
\end{table}

\begin{table}[tb]
\centering
\caption{V/UV recall under additive noise. REAPER is grayed out because it was used to compute the speech ground truth.}
\vspace{5pt}
\label{tb:noise_recall}
\renewcommand{\arraystretch}{1}
\resizebox{\columnwidth}{!}{
\begin{tabular}{lccccc}
\hline
\multirow{2}{*}{Method $\backslash$ SNR [dB]} & \multicolumn{5}{c}{V/UV recall$ \uparrow$} \\
\cmidrule(lr){2-6}
& $\infty$ & 30 & 20 & 10 & 0 \\
\hline
\color{gray}{REAPER}  & \color{gray}{91.68} & \color{gray}{89.87} & \color{gray}{83.21} & \color{gray}{63.21} & \color{gray}{50.31} \\
\cmidrule[0.05mm](lr){1-6}
RAPT    & 91.65 & 91.97 & 91.79 & 87.07 & 64.40 \\
SWIPE'  & 72.94 & 72.04 & 71.21 & 65.39 & 38.49 \\
pYIN    & 67.86 & 62.26 & 91.23 & 85.31 & 32.12 \\
DIO     & 78.16 & 72.44 & 69.08 & 77.68 & 46.87 \\
Harvest & 90.48 & 89.79 & 86.35 & 85.94 & 49.53 \\
Praat   & 84.94 & 83.96 & 83.83 & 81.58 & 54.35 \\
CREPE   & 87.98 & 84.93 & 84.27 & 80.89 & \textbf{64.86} \\
FCNF0++ & 89.92 & 89.02 & 86.85 & 82.41 & 63.81 \\
\cmidrule[0.05mm](lr){1-6}
Voting  & \textbf{94.21} & \textbf{92.90} & \textbf{91.40} & \textbf{89.50} & 52.27 \\
\hline
\end{tabular}
}
\vspace{6pt}
\end{table}

\begin{table}[tb]
\centering
\caption{Method sets selected by the accuracy criterion and by the correlation criterion.}
\vspace{5pt}
\label{tb:selection}
\renewcommand{\arraystretch}{1}
\resizebox{\columnwidth}{!}{
\begin{tabular}{llccc}
\hline
Criterion & Number & Set & $\mathrm{RPA}_{50}\uparrow$ & V/UV recall$\uparrow$ \\
\hline
--- & All & --- & 76.78 & 94.21 \\
\cmidrule[0.05mm](lr){1-5}
\multirow{3}{*}{Accuracy} & 3  & REAPER, RAPT, Harvest & 71.44 & 91.49 \\
& 5    & \begin{tabular}{c} REAPER, RAPT, DIO, \\Harvest, FCNF0++ \end{tabular}& 73.78 & 91.49 \\
\cmidrule[0.05mm](lr){1-5}
\multirow{3}{*}{Correlation}& 3  & REAPER, RAPT, FCNF0++ & 69.44 & 89.49 \\
& 5    & \begin{tabular}{c}REAPER, RAPT, Harvest, \\CREPE, FCNF0++ \end{tabular} & 71.74 & 92.39 \\
\hline
\end{tabular}
}
\end{table}

\section{Experimental evaluation}
\subsection{Experimental conditions}
We compared estimation errors, noise robustness, and the quality of analysis-by-synthesis audio across $f_\mathrm{o}$ estimators.
As component methods for voting and as baselines, we used
\textbf{RAPT}~\cite{talkin1995rapt},
\textbf{SWIPE'}~\cite{Camacho2008swipe},
\textbf{pYIN}~\cite{Mauch2014pyin},
\textbf{DIO}~\cite{morise2009dio},
\textbf{REAPER}~\cite{talkin2015reaper},
\textbf{Harvest}~\cite{morise2017harvest},
\textbf{Praat}~\cite{boersma2001praat},
\textbf{CREPE}~\cite{Kim2018crepe}, and
\textbf{FCNF0++}~\cite{morrison2023penn}.
We chose these methods considering both accuracy and implementation availability.
For the proposed \textbf{Voting} method, \textcolor{black}{we prepared three settings, with and without the alignment-based correction in \Subsec{alignment}.}
We also ran the method-selection algorithm in \Subsec{selection} with the accuracy and correlation criteria, constructing combinations with three and five methods.
We used \textbf{REAPER} as the initial element of the greedy algorithm.

For datasets, we used resampled audio at 48~kHz and 16~bit from
speech (Bagshaw~\cite{bagshaw1993pitch}, Keele~\cite{plante1995keele}, CMU ARCTIC~\cite{kominek2004cmu-arctic}, PTDB-TUG~\cite{pirker2011ptdb-tug}, MOCHA-TIMIT~\cite{wrench1999mocha-timit}),
singing voice (MIR-1K~\cite{hsu2010mir-1k}), and
instrumental sounds (MDB-stem-synth~\cite{salamon2017mdb-stem-synth}).
For datasets that provide only the electroglottograph (EGG) waveform, we applied noise reduction~\cite{rx11} and then computed the ground truth using \textbf{REAPER}, which performs well for EGG analysis.
To compute errors when the analysis times differed by less than one sample between the ground truth and an estimate, we linearly interpolated the ground-truth sequence.
We set the lower and upper bounds of $f_\mathrm{o}$ to 25~Hz and 4200~Hz, respectively, and we used a frame shift of 5~ms for all methods.

As error measures, we used the mean and the standard deviation of the pitch error in cents $\Delta \cent$,
RPA, and the average of V/UV errors.
For RPA, we used three thresholds, 5, 25, and 50~cents, where 5 is close to the just noticeable difference of pitch and 50 corresponds to a semitone.
For V/UV evaluation, we treated voiced frames as positives and unvoiced frames as negatives and computed recall and the false alarm rate.
High recall is desirable when $f_\mathrm{o}$ is used for speech synthesis, while a small false alarm rate is desirable when it is used for melody estimation.

We also evaluated the quality of analysis-by-synthesis speech using the estimated $f_\mathrm{o}$.
When the estimates contain errors, the quality of the synthesized speech is expected to degrade.
We used WORLD~\cite{Morise2016world} (D4C edition~\cite{morise2016d4c}) to extract the spectral envelope and aperiodicity measures and to resynthesize speech.
We used mean opinion score (MOS) for naturalness, and we estimated it automatically using UTMOSv2~\cite{baba2024utmosv2} as a proxy for subjective evaluation.

To evaluate robustness to noise, we added background noise samples chosen at random from NOISEX92~\cite{noisex92} and QUT-NOISE~\cite{dean2010qutnoise} at signal-to-noise ratios (SNRs) of
30, 20, and 10~dB and computed RPA and V/UV recall.
We also report the case with no additive noise as $\infty$~dB.

\subsection{Results and discussion}
\Table{no_noise} shows the results in the clean condition without additive noise.
Excluding \textbf{REAPER}, which we used to compute the ground truth,
\textbf{pYIN} achieved the best false alarm rate.
\textbf{pYIN} estimates periodicity by autocorrelation of the time-domain waveform, so it is less likely to misinterpret weak noise or unvoiced consonants as harmonics.
For the other measures, the voting method achieved the best accuracy.
In particular, the voting method outperformed REAPER in V/UV recall.
This suggests that voting is robust across domains, including music and singing voice.
\textcolor{black}{The bias correction by temporal and frequential alignment was also effective, as both improved the scores.}

\Table{noise_rpa} and \Table{noise_recall} show RPA and V/UV recall under additive noise.
As SNR degrades, in general, both measures drop markedly.
\textcolor{black}{Under noisy conditions, the voting method shows increasing errors in the $f_\mathrm{o}$ trajectories as SNR decreases, while maintaining relatively high V/UV recall down to low SNRs.
However, in extremely noisy environments, its performance eventually degrades and becomes inferior to single DNN-based estimators such as \textbf{CREPE} and \textbf{FCNF0++}.}
It is generally reported that DNN-based methods are robust to noise~\cite{Kim2018crepe, terashima2025slash}, and our results reflect this tendency.

\Table{selection} shows the results of limiting the number of methods and running the selection algorithm.
With both criteria, accuracy improved as the number of methods increased.
Despite the different criteria, the selected combinations were similar.
This observation supports the assumptions in \Subsec{variance}, and it suggests that one can construct a high-performing set only from the error correlations among methods even without access to ground truth.

\section{Conclusion}
In this paper, we theoretically analyze the voting method for fundamental frequency estimation and propose two improvements to enhance its performance: 1) a pre-alignment procedure to correct temporal and frequential biases among estimators, and 2) a greedy algorithm to efficiently select a compact, effective subset of estimators based on error correlation.

Experiments on the datasets speech, singing voice and music showed that the proposed method achieves higher accuracy than individual state-of-the-art estimators under clean conditions and enables robust voiced/unvoiced detection even in noisy environments. We confirmed, in particular, that the proposed alignment procedure is essential for improving accuracy.

Future work includes improving the estimation accuracy of the frequency trajectory under noisy conditions. Future work includes exploring combinations with a wider variety of methods and investigating robustness to in-the-wild signals.
\vfill\pagebreak
\bibliographystyle{IEEEbib}
\bibliography{bibtex/icassp25}

\end{document}